\documentclass[12pt]{article}
\usepackage{amssymb}
\usepackage{epsfig}

\newcommand{\bq}{\begin{equation}}
\newcommand{\eq}{\end{equation}}
\newcommand{\bqa}{\begin{eqnarray}}
\newcommand{\eqa}{\end{eqnarray}}
\newcommand{\ben}{\begin{enumerate}}
\newcommand{\een}{\end{enumerate}}
\newcommand{\bc}{\begin{center}}
\newcommand{\ec}{\end{center}}

\def\lsim{\lesssim}

\def\pr#1#2#3{ Phys. Rev. ${\bf{#1}}$ (#2) #3}

\def\pl#1#2#3{ Phys. Lett. ${\bf{#1}}$ (#2) #3}

\def\np#1#2#3{ Nucl. Phys. ${\bf{#1}}$ (#2) #3}
\def\zp#1#2#3{ Z. f. Phys. ${\bf{#1}}$ (#2) #3}

\global\nulldelimiterspace = 0pt

\def\O{ {\cal O }}

\begin{document}
\thispagestyle{empty}
\begin {flushleft}

 PM/97-30\\

August 1997\\
\end{flushleft}

\vspace*{2cm}

\hspace*{-0.5cm}
\begin{center}
{\Large {\bf Signals for Neutralino Box Effects at LEP2}}\\

\vspace{1.cm}
\centerline{\Large{J. Layssac$^{\rm a}$, F.M. Renard$^{\rm a}$,and 
            C. Verzegnassi$^{\rm b}$}}

\vspace {0.5cm} 
\centerline{\small  $^a$ \it Physique
Math\'{e}matique et Th\'{e}orique, UPRES-A 5032,}
 \centerline{\small \it Universit\'{e} Montpellier
II, F-34095 Montpellier Cedex 5.}
\vskip 0.3cm
\centerline{\small $^{\rm b}$\it Dipartimento di Fisica, Universit\`a di 
                   Lecce and INFN, Sezione di Lecce,}
\centerline{\small \it via Arnesano, 73100 Lecce, Italy}

\vspace*{3cm}
{\bf Abstract}\hspace{2.2cm}\null
\end{center}
\hspace*{-1.2cm}
\begin{minipage}[b]{16cm}
We have computed the contribution to the observables of the final
two fermion channel at LEP2, at the limiting energy
$\sqrt{q^2}=200~GeV$, coming from boxes with two neutralinos of
purely gaugino type, of mass $M=100~GeV$. We find a potentially
visible effect only for the muon channel, in the cross section and,
to a lesser extent, in the forward-backward asymmetry. Analogous
effects coming from the chargino box are also briefly discussed.\\

\vspace{3cm}

\end{minipage} 

\setcounter{footnote}{0} 
\clearpage
\newpage 
  
\hoffset=-1.46truecm
\voffset=-2.8truecm
\textwidth 16cm
\textheight 22cm
\setlength{\topmargin}{1.5cm}

\hspace{0.7cm}Among the various proposed theoretical interpretations
\cite{R1} of the four-jet events excess at LEP2 reported by the ALEPH
collaboration \cite{R2}, the supersymmetric mechanism with R-parity
violation suggested by Carena, Giudice, Lola and Wagner \cite{R3} has
been recently considered with special attention. This is not only due
to the several intrinsic virtues of the proposal, that seems to be able
to explain remarkably well the most characteristic features of the
ALEPH data, but also to the fact that the same mechanism would provide
a satisfactory interpretation \cite{R4}
of the excess of large $Q^2$ events of
neutral current type (positron in the final state) recently observed at
HERA \cite{R5}. This is made possible by the fact that one restriction
on the model coming from LEP2 is that the mass of the neutralino
exchanged in the t-channel must lie in a range between $\simeq$ 80 and
100 GeV (owing to the kind of diagram involved, only a neutralino of
purely gaugino type -bino, wino- would be involved). This restriction
is not inconsistent with what is requested in order to explain the HERA
neutral current excess.\par
If the theoretical mechanism proposed in ref.\cite{R3} corresponds to
physical reality, several possible visible consequences would be
effective in a near future at LEP2. In particular, direct production of
couples of sleptons and/or of a couple of neutralinos should become
detectable. Correspondingly, new excesses in the final 4 jets channel
should be seen in the first case. For neutralinos, direct production
might be less evident, particularly for values of the mass close to the
upper suggested value $M_{\chi^0} \simeq 100~GeV$. In case of future
evidence of the proposed slepton production, it would be highly
welcome, for a self-consistent test of the overall picture, to identify
the presence of the suitable neutralino even if its mass lied in the
unfavored region around approximately 100 GeV.\par
The aim of this paper is precisely that of showing that, should the
proposed neutralino mass lie indeed in the 100 GeV region, it would
still be possible to predict and detect a sizeable signal in the final
\underline{muon} channel (or, more generally, in the final "lepton" to
be suitably defined e.g. by considering muon and $\tau$
production-channel). This would be due to a rather special virtual
one-loop effect, exclusively produced by neutralino boxes.\par
A few words of comments are at this point appropriate. For what
concerns LEP1 physics, the one-loop virtual electroweak box
contribution is systematically negligible in the theoretical
expression of the various observables, to the extent that such
contributions are meant to be computed at
$q^2=(p_{e^-}+p_{e^+})^2=M^2_Z$ (this statement does not apply e.g. to
the box contributions to the redefinition of $G_{\mu}$, where they are
computed at $q^2=0$). This can be qualitatively understood since, in
the various observables, certain gauge-invariant combinations of
self-energies, vertices and boxes appear at the one loop level,
whose box component carries a multiplicative factor $\simeq
(q^2-M^2_Z)$ that vanishes exactly on Z resonance. When one moves
away from the peak, this feature is completely reversed. In
particular, the naive expectation is that, when $q^2$ increases, the
relative weight of box contributions becomes enhanced until it
reaches a "stable" regime at sufficiently large $q^2$ values. One
can define this feature as an expected "kinematical" box
enhancement. Note that, strictly speaking, these guesses are
supposed to be valid for a final two fermion state, for which all
variables can be continuously continuated from the Z peak to higher
$q^2$ values. In fact, from a glance at the existing rigorous SM
calculations at one-loop \cite{R6}, one verifies that, indeed, this
expectation is verified and in particular that, in the LEP2 energy
range $\sqrt{q^2}\lsim 200~GeV$, a sizeable and evidently
"kinematical" increase shows up when one approaches the limiting
value $\sqrt{q^2}\simeq 200~GeV$.\par
The previous qualitative considerations can be made technically more
plausible if one adopts for the final two fermion processes in
electron-positron annihilation a theoretical description defined as
"Z-peak subtracted" representation \cite{Zsub1},\cite{Zsub2}. In such an
approach, a kinematical box enhancement $\simeq(q^2-M^2_Z)$ appears as
the logical consequence of the fact that for the remaining one loop
quantities (self-energy, vertices) a systematical subtraction procedure
can be performed that makes their contribution, for models of
electroweak type, intrinsically depressed with respect to the boxes'
one. We do not insist on this point here, since the presentation of
refs.\cite{Zsub1},\cite{Zsub2} is sufficiently detailed, and defer to a 
forthcoming paper for a longer and systematic discussion about boxes'
relevance.\par
So finally, on a purely kinematical basis, one would expect that in
a supersymmetric model like the MSSM whose analogies with the MSM are
often remarkable, an enhancement of these boxes that correspond to the
MSM ones (with WW and ZZ s-channel exchange) appears when moving
towards the highest c.m. energy values (in our work, assumed for
simplicity to be at $\sqrt{q^2}\simeq 200~GeV$). This should be valid,
in particular, for neutralino boxes on which our attention is now
concentrated.\par
A peculiar feature of the supersymmetric scenario should be now
stressed. In the MSM, the relative
importance of ZZ boxes is much smaller than that of WW boxes, and one
might feel that the same feature should remain in the MSSM. As a matter
of fact, the situation here \underline{might} be rather different,
since the genuinely electroweak contributions now arise, not only from
a "zino", but also from a "photino" contribution, while photon boxes
were not included in the electroweak sector in the SM, but in the QED
corrections. Thus, a priori, neutralino boxes might be relevant in
the MSSM.\par
On top of the previously hypothized kinematical enhancement, boxes can
exhibit one extra type of enhancement of
"dynamical" type, corresponding to a sort of threshold effect that
shows up when $\sqrt{q^2}$ approaches the value $M_1+M_2$, where
$M_1,M_2$ are the masses of the two particles that are exchanged in the
box (in principle, they might be different). This enhancement is not
peculiar of box diagrams, and would appear in self-energies and
vertices as well. In the SM specific case, one can actually see such an
enhancement e.g. at $\sqrt{q^2}=2M_W$, and verify that around the
threshold value a "sizeable" (typically, of a relative one-two percent)
effect is produced, at least in certain observables. Clearly, this
possible enhancement is only fixed by the relevant particle's masses,
and is independent of $\sqrt{q^2}$.\par
The simple observation on which our paper is based is that there exist,
in principle, situations in which these two boxes' enhancement effects
would sum up. They correspond to the rather priviledged case in which
the highest energy that is available corresponds, at least
approximately, to the sum of the particles' masses. This would be
exactly the case of a box with two neutralinos of the type suggested in
ref.\cite{R3}, both with a mass of 100 GeV, at an energy
$\sqrt{q^2}\simeq200~GeV$ (that represents a possible goal for LEP2).
Note that, for the supposed combination of purely gaugino content, no
virtual contribution from self-energies or double neutralino
vertices would be allowed.
Since direct production in this mass-energy configuration would not be
feasible, the box with two neutralinos would represent in this case the
only possibly visible effect due to such particles in the final two
fermion state. We have consequently computed the related contribution,
and will devote the following part of the paper to a discussion of the
numerical analysis and results.\par
The neutralino effect via boxes corresponds to the two diagrams
depicted in Fig.(1). We have computed the contribution to the
invariant scattering amplitude for a final fermion-antifermion
state.
 Our analysis, as well as that of
ref\cite{R6}, will be systematically performed in the 't Hooft gauge
$\xi=1$. Final results are obviously gauge-independent.\par

The exchanged sfermions are $\tilde{e}_{L,R}$ and $\tilde{f}_{L,R}$ 
(for simplicity we will take $L$ and $R$ states
with a common mass; in the numerical applications
we have taken $m_{\tilde{l}}=60~GeV$ and
$m_{\tilde{q}}=110~GeV$). The intermediate neutralinos ($i$ or $j =
1,..4$) in principle
consist of four independent Majorana states constructed as 
mixtures of pure gaugino
($\tilde{W}$,$\tilde{B}$) and pure higgsino
($\tilde{H}_1$,$\tilde{H}_2$) types. The higgsino components couple to 
($f\tilde{f}$) proportionally to the fermion mass and 
we will neglect them.
The gaugino couplings to ($f\tilde{f}$) are listed in Table 1. 
The complete
amplitude is obtained by summing all 64 combinations
($\tilde{e}_{L,R}$,$\tilde{f}_{L,R}$,$\chi^0_i$,$\chi^0_j$).
For simplicity we will take all neutralinos
with a common mass $M=100~ GeV$. In this case the sum over all intermediate
combinations can be expressed in terms of the pure gaugino contributions
($\tilde{W}\tilde{W}$), ($\tilde{B}\tilde{B}$), ($\tilde{W}\tilde{B}$), 
($\tilde{B}\tilde{W}$). After standard but lengthy Dirac 
algebra the total box
amplitude of the $e^+e^- \to f\bar f$ process can be written as

\bqa
A^{NB}(e^+e^-\to f\bar f)&=&{g^4\over16c^4_W}I_N \{H_{LL} L^{\mu}_{ee}
L_{\mu,ff}+16s^4_W H_{RR} R^{\mu}_{ee}R_{\mu,ff}\nonumber\\
&&~~~~~~~~~~+4s^2_W H_{LR} L^{\mu}_{ee}R_{\mu,ff}+4s^2_W H_{RL} R^{\mu}_{ee}
L_{\mu,ff}\}
\eqa
\bq
H_{XY}=(h^{\tilde{W}}_{Xe})^2 (h^{\tilde{W}}_{Yf})^2 
+(h^{\tilde{B}}_{Xe})^2 (h^{\tilde{B}}_{Yf})^2
+2(h^{\tilde{W}}_{Xe})(h^{\tilde{W}}_{Yf})
(h^{\tilde{B}}_{Xe})(h^{\tilde{B}}_{Yf})
\eq
\noindent
with $X$ or $Y=L,R$ and $h^{\tilde{W},\tilde{B}}_{L,R f}$ given in
Table 1; also
\bq
R^{\mu}_{ee},L^{\mu}_{ee}=\bar v(e^+)\gamma^{\mu}(1\pm\gamma^5)u(e^-)\
\ \ \ \ \ 
R^{\mu}_{ff},L^{\mu}_{ff}=\bar u(f)\gamma^{\mu}(1\pm\gamma^5)v(\bar f)
\eq
\noindent
and $I_N$ is a combination of Feynman box integrals computed numerically
through the Passarino-Veltman method, ref.\cite{Rint}.

To compute the interference effect with the SM
$e^+e^-\to\gamma,Z\to f\bar f$ amplitude
within the Z-peak subtracted method, 
it is
convenient to decompose the expression in eq.(1) on photon and Z Lorentz
structures as given in ref.\cite{Zsub1},\cite{Zsub2}:
\bqa
A^{NB}(e^+e^-\to f\bar f)&=&v_{ l}^{\mu(\gamma)}v_{\mu f}^{(\gamma)}
A^{NB}_{\gamma\gamma,lf}(q^2,cos\theta)~
+~v_{ l}^{\mu(Z)}v_{\mu f}^{(Z)}A^{NB}_{ZZ,lf}(q^2,cos\theta)\nonumber\\
&&+~v_{ l}^{\mu(\gamma)}v_{\mu f}^{(Z)}
A^{NB}_{\gamma Z,lf}(q^2,cos\theta)~
+~v_{ l}^{\mu(Z)}v_{\mu f}^{(\gamma)}A^{NB}_{Z\gamma,lf}(q^2,cos\theta)
\eqa
\noindent
with

 \bq v_{\mu f}^{(\gamma)}=
eQ_f\bar u(f)\gamma_{\mu}v(\bar f) \ \ \ \ \ \ \  v_{\mu f}^{(Z)}=
{g\over c_w}
\bar u(f)\gamma_{\mu}(\tilde{v}_f-2\gamma^5 I_{3f})v(\bar f)
\eq
\noindent
and
\bqa
A^{NB}_{\gamma\gamma,lf}(q^2,cos\theta)&=&
{g^4\over16e^2 c^4_W Q_e Q_f}I_N \{
H_{LL}(1-\tilde{v}_e)(1-\tilde{v}_f)
+16s^4_W H_{RR}(1+\tilde{v}_e)(1+\tilde{v}_f)\nonumber\\
&&~~~~+4s^2_W
[H_{LR}(1+\tilde{v}_e)(1-\tilde{v}_f)
+H_{RL}(1-\tilde{v}_e)(1+\tilde{v}_f]\}
\eqa

\bqa
A^{NB}_{ZZ,lf}(q^2,cos\theta)&=&
{g^4s^2_W\over4e^2 c^2_W I_{3e} I_{3f}}I_N \{
H_{LL}-16s^4_W H_{RR}-4s^2_W [H_{LR}+H_{RL}]\}
\eqa
\bqa
A^{NB}_{\gamma Z,lf}(q^2,cos\theta)&=&
{g^4s_W\over8e^2 c^3_W Q_{e} I_{3f}}I_N \{
H_{LL}(1-\tilde{v}_e)-16s^4_W H_{RR}(1+\tilde{v}_e)\nonumber\\
&&+4s^2_W [H_{LR}(1+\tilde{v}_e)-H_{RL}(1-\tilde{v}_e)]\}
\eqa
\bqa
A^{NB}_{Z\gamma,lf}(q^2,cos\theta)&=&
{g^4s_W\over8e^2 c^3_W I_{3e} Q_{f}}I_N \{
H_{LL}(1-\tilde{v}_f)-16s^4_W H_{RR}(1+\tilde{v}_f)\nonumber\\
&&-4s^2_W [H_{LR}(1-\tilde{v}_f)-H_{RL}(1+\tilde{v}_f)]\}
\eqa 
\noindent

\noindent
with
 $\tilde{v}_f\equiv 1-4|Q_f|s^2_f$ defined with $s^2_f$, 
the effective Weinberg
angle for the $f$-fermion,\cite{Zsub2}.
 
The contribution to the various observables is then immediately
obtained (see the explicit derivation in ref.\cite{Zsub2})
in terms of the four quantities
\bq
\tilde{\Delta}_\alpha^{(NB,lf)}(q^2,cos\theta)=q^2
A^{(NB,lf)}_{\gamma\gamma,lf}(q^2,cos\theta)
\label{Da}\eq
\bq
R^{(NB,lf)}(q^2,cos\theta)=-(q^2-M^2_Z) A^{NB}_{ZZ,lf}(q^2,cos\theta)
\label{R}\eq
\bq
V^{(NB,lf)}_{\gamma Z}(q^2,cos\theta)=
-(q^2-M^2_Z) A^{NB}_{\gamma Z,lf}(q^2,cos\theta)
\label{VgZ}\eq
\bq
V^{(NB,lf)}_{ Z\gamma}(q^2,cos\theta)
=-(q^2-M^2_Z) A^{NB}_{Z\gamma,lf}(q^2,cos\theta) \ \ .
\label{VZg}\eq

Starting from the two parts $\sigma^{lf}_{1,2}(q^2,cos\theta)$
of the angular distribution (expressed in terms of Z-peak inputs
\cite{Zsub1},\cite{Zsub2})
\bqa
&&\sigma^{lf}_1(q^2,cos\theta)=N_f{4\pi
q^2\over3}\{\alpha^2(0)Q^2_f[1+2\tilde{\Delta}^{(lf)}
_\alpha(q^2,cos\theta)]
\nonumber\\
&&+2[\alpha(0)Q_f]{q^2-M^2_Z\over
q^2((q^2-M^2_Z)^2+M^2_Z\Gamma^2_Z)}[{3\Gamma_l\over
M_Z}]^{1/2}[{3\Gamma_b\over N_f M_Z}]^{1/2}
{\tilde{v}_l \tilde{v}_f\over
(1+\tilde{v}^2_l)^{1/2}(1+\tilde{v}^2_f)^{1/2}}\nonumber\\
&&\times[1+
\tilde{\Delta}^{(lf)}_\alpha(q^2,cos\theta) 
-R^{(lf)}(q^2,cos\theta)
-4s_lc_l
\{{1\over \tilde{v}_l}V^{(lf)}_{\gamma Z}(q^2,cos\theta)
+{1\over 3\tilde{v}_f}
V^{(lf)}_{Z\gamma}(q^2,cos\theta)\}]\nonumber\\ 
&&+{[{3\Gamma_l\over
M_Z}][{3\Gamma_f\over N_f M_Z}]\over(q^2-M^2_Z)^2+M^2_Z\Gamma^2_Z}
\times[1-2R^{(lf)}(q^2,cos\theta)
-8s_lc_l\{{\tilde{v}_l\over1+\tilde{v}^2_l}V^{(lf)}_{\gamma
Z}(q^2,cos\theta)\nonumber\\
&&+{\tilde{v}_f\over3(1+\tilde{v}^2_f)}
V^{(lf)}_{Z\gamma}(q^2,cos\theta)\}]\}\ , 
\label{sigma1}
\eqa

\bqa
&&\sigma^{lf}_2(q^2,cos\theta)={3\over4}N_f{4\pi
q^2\over3}\{ 2[\alpha(0)Q_f]{q^2-M^2_Z\over
q^2((q^2-M^2_Z)^2+M^2_Z\Gamma^2_Z)}
[{3\Gamma_l\over M_Z}]^{1/2}[{3\Gamma_f\over N_f
M_Z}]^{1/2}\nonumber\\
&&{1\over(1+\tilde{v}^2_l)^{1/2}(1+\tilde{v}^2_f)^{1/2}}
[1+
\tilde{\Delta}^{(lf)}_\alpha(q^2,cos\theta) 
-R^{(lf)}(q^2,cos\theta)]\nonumber\\
&&+{[{3\Gamma_l\over
M_Z}][{3\Gamma_f\over N_f
M_Z}]\over(q^2-M^2_Z)^2+M^2_Z\Gamma^2_Z}
[{4\tilde{v}_l \tilde{v}_f\over(1+\tilde{v}^2_l)(1+\tilde{v}^2_f)}]
[1-2R^{(lf)}(q^2,cos\theta)\nonumber\\
&&-4s_lc_l
\{{1\over \tilde{v}_l}V^{(lf)}_{\gamma Z}(q^2,cos\theta)
+{1\over 3\tilde{v}_f}
V^{(lf)}_{Z\gamma}(q^2,cos\theta)\}]\} \ , 
\label{sigma2} 
\eqa
\noindent
one obtains the integrated cross section and the forward-backward
asymmetry as:
\bq
\sigma^{lf}=\int^{+1}_{1} dcos\theta[{3\over8}(1+cos^2\theta)
\sigma^{lf}_1+cos\theta\sigma^{lf}_2]
\eq

\bq
\sigma^{lf}_{FB}=[\int^{+1}_{0}-\int^{0}_{-1}] dcos\theta[{3\over8}(1+cos^2\theta)
\sigma^{lf}_1+cos\theta\sigma^{lf}_2]
\eq

\bq
A_{FB,lf}={\sigma^{lf}_{FB}\over\sigma^{lf}}
\eq

We now consider the observables which are measurable with the
highest accuracy at LEP2, namely
$\sigma^{\mu}$, the total cross section for muon pair production,
$A_{FB,\mu}$, its forward-backward asymmetry and $\sigma^5$,
the total hadron production. For each of them we compute the 
relative neutralino box effect by inserting in eq.(\ref{sigma1},
\ref{sigma2}) the contributions
of eq.(\ref{Da}-\ref{VZg}). For each observable this gives the relative
effect

\bq
{\delta\O\over\O} = {\O^{(SM+NB)}-\O^{(SM)}\over\O^{(SM)}}
\eq
 
We first discuss muon pair production. As expected the box effect peaks
at $\sqrt{s}=2M=200~GeV$ as one can see in Fig.2. At this energy
$\tilde{W}$ and $\tilde{B}$ contributions are of comparable 
magnitude and
cumulate to a total effect of about 1.4 percent on
$\sigma^{\mu}$. This is at an observable level at LEP2 \cite{LEP2} at
the optimal expected experimental accuracy of about relative 0.7
percent \cite{LEP2} for $\sigma^{\mu}$ (and $A_{FB,\mu}$). The
forward-backward asymmetry gets also an effect which peaks at 200 GeV
but it is relatively weaker (relative 0.7 percent which means 0.4
percent absolute on $A_{FB,\mu}$) at the limiting observability.\par
We have also computed the effects on quark pair production. As one can
expect from the weaker neutralino couplings given in Table 1, 
the separate effects on $u\bar u$ and $d\bar d$ are somewhat weaker
than on muon pair. But they have also an opposite sign for $u\bar u$ and
for $d\bar d$ so that they largely cancel in $\sigma^5$, leaving only a
peak of -1 permille. So finally this neutralino box effect
consists in a positive effect on the muon pair cross section,
correlated to a negative effect on the forward-backward asymmetry and no
effect on total hadron production. It peaks at $\sqrt{s}=2M$,
 with a kinematical half-width
of about 15 GeV ($2M\pm15~GeV$).\par
For comparison and check we have also looked at the chargino case.
Assuming that two degenerate couples $\chi^{\pm}_i$ exist with a mass
$M=100~GeV$, we have computed the corresponding box effect around
$\sqrt{s}=200~GeV$ (in fact, for values $\sqrt{s}<200~GeV$ where
direct production is not possible). In this case there is only one box 
diagram with ($\nu_e,\tilde{f'}$) exchange and 
intermediate $\chi^+_i\chi^-_j$. Summing over all
degenerate $\chi_i$ states or just taking one couple of pure gaugino
$\tilde{W}^{\pm}$, with pure left
coupling to $ f \tilde{f'}$, one gets a box effect 
which peaks at $\sqrt{s}=2M$ with a magnitude of
2 percent on $\sigma^{\mu}$, 5 permille on $A^{\mu}_{FB}$ and 1
percent on $\sigma^5$. This effect is very similar but opposite in sign
to the standard one due to the WW box. In fact we have checked that
apart from mass differences the projection on photon and Z Lorentz
structures of the WW and pure gaugino $\chi\chi$ boxes have the same
leading expressions. This chargino peak is therefore comparable to the
neutralino one (apart from a different sign on $A^{\mu}_{FB}$ and a
larger effect on $\sigma^5$). Remember also that there are now
additional chargino
contributions (self-energy, vertices) which decrease with the energy
relatively to the box contributions as explained in the introdution.\par

In fact, we have also computed the overall, gauge invariant combination
of chargino self-energy, vertices and boxes and found that, within the
combination, the box contribution remains the dominant one, in
agreement with our general and previously discussed expectations based
on the theoretical subtracted approach that we have used.\par

In conclusion,
motivated by an interesting theoretical suggestion whose experimental
confirmation is still debated, we have verified that, in a conventional
minimal supersymmetric extension of the Standard Model, certain virtual
one-loop contributions of box type might have visible effects in the
simple and clean final (two) lepton channel at LEP2. The origins of
this fact are partially due to a special kinematical enhancement
property of box effects, that might make them specially relevant in
general at increasing c.m. energy in an electron-positron collision.
This should remain valid even in cases where the extra "dynamical"
enhancement produced by a "quasi" resonant configuration were absent.
In this case, we would have at $\sqrt{s}>M_Z$ a situation in which for
virtual effects there might be a kind of "box dominance", quite
orthogonal to the situation met on top of $Z$ resonance, that would be
more effective when the c.m. energy increases (typically this might be
quite relevant for a future 500 GeV LC collider). A systematic
investigation along this line is by now in progress.

{\bf \underline{Acknowledgments}}\par
We thank Gilbert Moultaka for having helped us in the use of the
Passarino-Veltman method and the van Oldenborgh package.
\newpage

\vspace{1cm}
\begin{center}
{\bf Table 1: Gaugino couplings to fermion-sfermion pairs.}

\vspace{1cm}
\begin{tabular}{|c|c|c|c|c|} \hline
\multicolumn{1}{|c|}{}&
\multicolumn{1}{|c|}{$h_{Le}$} &
\multicolumn{1}{|c|}{$h_{Re}$}&
\multicolumn{1}{|c|}{$h_{Lq}$}&
\multicolumn{1}{|c|}{$h_{Rq}$} 
 \\[0.1cm] \hline
&&&&\\
$\tilde{B}$&$s_W$&$-1$&$-{1\over3}s_W$&$Q_q$\\
&&&&\\
$\tilde{W}$&$c_W$&$0$&$-2I_{3q}c_W$&$0$\\
&&&&\\
 \hline
\end{tabular}
\end{center}

\newpage

\begin{center}

{\large \bf Figure captions}
\end{center}
\vspace{0.5cm}

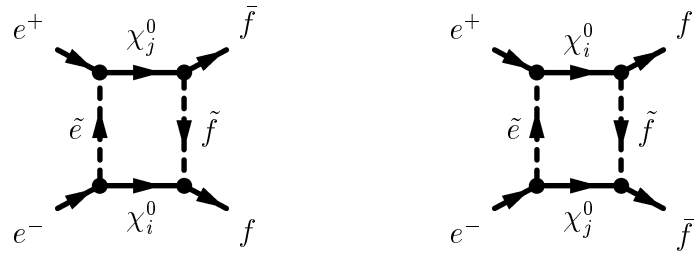
{\bf Fig.1} Box diagrams with sfermion exchanges and intermediate
neutralinos for the $e^+e^-\to f\bar f$ amplitude.\\

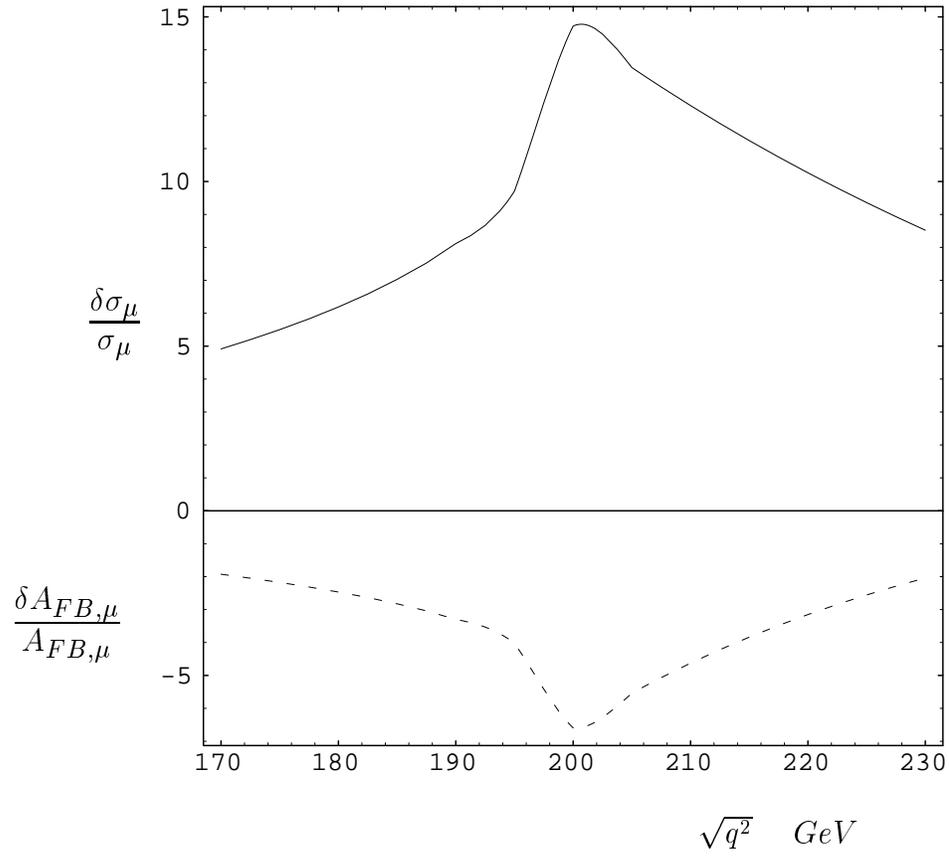
{\bf Fig.2} Neutralino box effect on muon pair production in the
LEP2 energy range. Relative effect on the cross section (solid);
on the forward-backward asymmetry(dashed).


\newpage


\begin{thebibliography}{99}


\bibitem{R1} D. Buskulic et al. (ALEPH Coll.),\zp{C71}{1996}{179}.
\bibitem{R2} F. Ragusa for the ALEPH Coll., talk at the LEPC Meeting,
November 19, 1996.
\bibitem{R3} M. Carena, G.F. Giudice, S. Lola, C.E.M. Wagner,
\pl{B395}{1997}{225}.
\bibitem{R4} See the discussion given by G. Altarelli, J. Ellis,
G.F. Giudice, S. Lola, M.L. Magano, hep-ph/9703276. 
\bibitem{R5} C Adloff et al., H1 collaboration, DESY preprint 97-24,
hep-ex/9702012.\\
J. Breitweg et al., ZEUS collaboration, DESY preprint
97-25, hep-ex/9702015.
\bibitem{R6} ``TOPAZ0"
G.~Montagna, O.~Nicrosini, G.~Passarino 
and F.~Piccinini,
``{\tt TOPAZ0~2.0} - A program for computing deconvoluted and realistic
observables around the $Z^0$ peak'', CERN-TH.7463/94; 
G.~Montagna, O.~Nicrosini, G.~Passarino, F.~Piccinini 
and R.~Pittau,
Comput. Phys. Commun. { 76} (1993) 328. 
\bibitem{Zsub1} F.M.~Renard and C.~Verzegnassi, \pr{D52} 
{1995}{1369}.
\bibitem{Zsub2} F.M.~Renard and C.~Verzegnassi,\pr{D53}{1996}{1290}.
\bibitem{Rint} G. Passarino and M.Veltman, \np {B160}{1979}{151}.\\
G.J. van Oldenborgh, FF-a package to evaluate one-loop Feynman diagrams,
Comp.Phys.Comm.66 (1991) 1.\\
Rolf Mertig, Guide to FeynCalc (1992) 
\bibitem{LEP2} See e.g. the discussion given in :
Physics at LEP2, Proceedings of the Workshop-Geneva,
Switzerland (1996), CERN 96-01, 
G. Altarelli, T. Sjostrand and F. Zwirner eds. 








\end{thebibliography}
\end{document}